\documentclass[journal=nalefd,manuscript=letter, layout=twocolumn]{achemso}

\usepackage[version=3]{mhchem} 
\usepackage{xcolor} 
\usepackage[colorinlistoftodos,prependcaption,textsize=tiny]{todonotes} 
\usepackage{cuted}
\usepackage{siunitx}

\newcommand{\figref}[2]{Fig.~\ref{#1}#2}      

\usepackage[labelfont=bf]{caption}

\makeatletter
\renewcommand*{\acs@author@fnsymbol@symbol}[1]{
    \ifcase #1 *\or
    1\or
    2\or
    3\or
    4\or
    5\or
    6\or
    7\or
    8\or
    9\or
    10
    \fi
}
\renewcommand*\acs@contact@details{
    {\sffamily *Corresponding Author: \protect\\ Hans Hilgenkamp (\acs@email@list)}%
    \acs@number@list
}   
\patchcmd{\acs@address@list@auxii}
{\acs@author@fnsymbol{\acs@affil@marker@cnt}}
{\textsuperscript{\acs@author@fnsymbol{\acs@affil@marker@cnt}}}
{}{}

\patchcmd{\acs@address@list@auxii}
{{\acs@author@fnsymbol{\acs@affil@marker@cnt}\@nameuse{@altaffil@\@roman\@tempcnta}\par}}
{{\textsuperscript{\acs@author@fnsymbol{\acs@affil@marker@cnt}}\@nameuse{@altaffil@\@roman\@tempcnta}\par}}
{}{}        

\makeatother 

\author{Thijs J. Roskamp}
\affiliation{\textnormal{Faculty of Science and Technology and MESA+ Institute, University of Twente, 7500 AE Enschede, The Netherlands}}

\author{Tim Horstink}
\affiliation{\textnormal{Bruker Nano Surfaces \& Metrology, De Veldmaat 17, 7522 NM Enschede, The Netherlands}}

\author{Melissa J. Goodwin}
\affiliation{\textnormal{Faculty of Science and Technology and MESA+ Institute, University of Twente, 7500 AE Enschede, The Netherlands}}

\author{Erwin Berenschot}
\affiliation{\textnormal{Faculty of Science and Technology and MESA+ Institute, University of Twente, 7500 AE Enschede, The Netherlands}}

\author{Edin Sarajilic}
\affiliation{\textnormal{Bruker Nano Surfaces \& Metrology, De Veldmaat 17, 7522 NM Enschede, The Netherlands}}

\author{Roeland Huijink}
\affiliation{\textnormal{Bruker Nano Surfaces \& Metrology, De Veldmaat 17, 7522 NM Enschede, The Netherlands}}

\author{Niels Tas}
\affiliation{\textnormal{Faculty of Science and Technology and MESA+ Institute, University of Twente, 7500 AE Enschede, The Netherlands}}

\author{Hans Hilgenkamp}
\affiliation{\textnormal{Faculty of Science and Technology and MESA+ Institute, University of Twente, 7500 AE Enschede, The Netherlands}}
\email{j.w.m.hilgenkamp@utwente.nl}

\title[]
  {Nanoscale wireframe SQUID on a cantilever by corner lithography}

\begin{document}

\begin{strip}
\vspace{-2cm}
\begin{abstract}
We present the fabrication of nanoscale superconducting quantum interference devices (SQUIDs) at the apex of wireframe tips on self-aligned superconducting cantilever probes. The probes are made on silicon wafers using molding techniques in combination with corner lithography, which results in a nanowire frame tip with a tuneable apex structure. A shadow effect deposition using magnetron sputtering of Nb creates self-aligned superconducting wireframes on cantilevers with accompanying device circuitry. Superconducting weak links are realized at the apex of the wireframes with the use of focused ion beam nanopatterning. The realized SQUIDs have effective diameters ranging from several micrometers down to 100 nm and can be operated in magnetic fields up to 1 T. Furthermore, the nanowires in the wireframe can be used to flux modulate the SQUID locally. This fabrication process enables the production of wafer-scale templates for probes based on on-tip superconducting devices.
\end{abstract}
\end{strip}
\begin{figure}[h]
    \centering
    \hspace*{3cm}\includegraphics[width=0.65\textwidth]{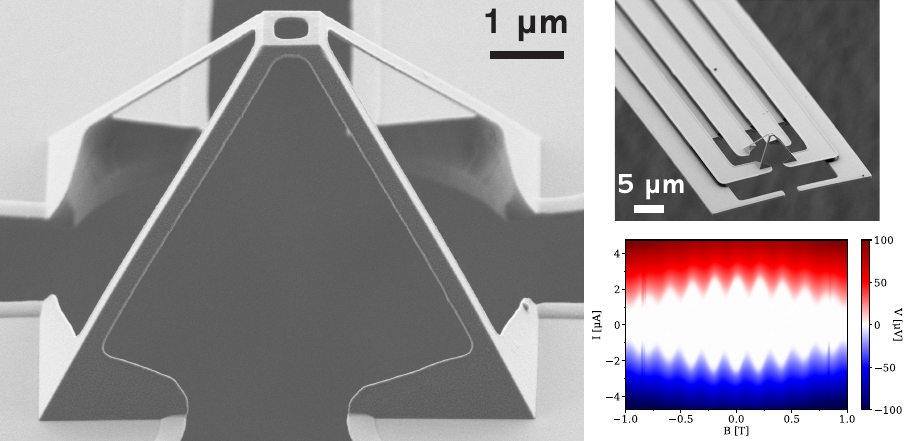}
    \label{fig:GraphAbstract}
\end{figure}
\clearpage
Superconducting quantum interference devices (SQUIDs) are the sensors with the highest magnetic field sensitivity \cite{kleiner_superconducting_2004, clarke_squid_2006, granata_nano_2016, martinez-perez_nanosquids_2017}. In scanning SQUID microscopy (SSM), SQUIDs are used for nanoscale magnetic imaging. Here, one uses a SQUID to locally map and image the magnetism from a sample \cite{kirtley_scanning_1999, kirtley_fundamental_2010, persky_studying_2022,christensen_2024_2024}. Among others, SSM has been used to image the order-parameter symmetry in high-T$_c$ superconductors\cite{kirtley_design_1995, kirtley_symmetry_1995, hilgenkamp_ordering_2003, kirtley_angle-resolved_2006}, the interplay between magnetism and superconductivity at oxide interfaces\cite{bert_direct_2011, persky_scanning_2018}, currents in topological insulators\cite{nowack_imaging_2013, spanton_images_2014, ferguson_direct_2023}, ferromagnetic domain patterns in magnetic oxides\cite{wang_imaging_2015, anahory_emergent_2016, folkers_imaging_2024}, Abrikosov vortices in type-II superconductors\cite{wells_analysis_2015,embon_probing_2015, bishop-van_horn_vortex_2023, keren_chip-integrated_2023}, and local dissipation in the quantum Hall state in graphene\cite{halbertal_nanoscale_2016, halbertal_imaging_2017}.\newline
In the past decades much effort has been dedicated to the development of SQUID probes for SSM, with the main focus to increase spatial resolution whilst maintaining high magnetic field sensitivity\cite{reith_analysing_2017, persky_studying_2022,christensen_2024_2024}. Traditionally, SQUID probes for SSM have been realized in large quantities on planar silicon wafers by conventional lithography\cite{kirtley_highresolution_1995, koshnick_terraced_2008, kirtley_scanning_2016}. These SQUID-on-chip probes have the advantage of additional on-chip circuitry such as flux modulation coils, a pickup loop, and coils to measure magnetic susceptibility, which together extend the functionalities of these SQUIDs. However, scanning in close proximity to the sample is hindered by the planar structure of the substrate chip and the spatial resolution is typically limited to only several micrometers\cite{persky_studying_2022, christensen_2024_2024}.\newline
To improve the spatial resolution of a SQUID probe, one has to bring the SQUID in closer proximity to the sample surface and reduce its effective area. This can be done by raising the SQUID onto an elevated surface like a sharp tip; the SQUID-on-tip probe was the first SQUID probe to achieve this\cite{finkler_self-aligned_2010, vasyukov_scanning_2013}. Here, a three-step thermal evaporation of a superconductor onto a pulled quartz capillary creates a self-aligned SQUID at the apex of a sharp tip. This method allows for the fabrication of the smallest SQUID sensors to date\cite{anahory_squid--tip_2020} and, by using conventional scanning probe feedback, for scanning within close proximity to the surface, increasing the spatial resolution to below 100 nanometers\cite{finkler_scanning_2012, zhou_scanning_2023}. However, the current SQUID-on-tip fabrication process hinders scalability; in addition the geometry of the quartz capillary precludes the addition of auxiliary circuitry. \newline
On the other hand, focused ion beam (FIB) milling on planar substrates has been an established technique to create nanoSQUIDS\cite{troeman_nanosquids_2007,hao_measurement_2008,schwarz_low-noise_2013, granata_three-dimensional_2013,anahory_three-junction_2014, potter_comparison_2025} and can be used in combination with commercial atomic force microscopy (AFM) cantilever probes to create SQUID probes\cite{wyss_magnetic_2022, Weber2025}. This fabrication method is highly versatile as it takes advantage of the precise nanoscale modification offered by FIB alongside the superior topographical feedback provided by AFM. However, utilizing commercial AFM probes presents significant challenge to scalable fabrication because of the requirement to individually prepare each probe by milling not only the SQUID, but also the the additional circuitry. \newline
Here, we combine wafer-scale molding and preparation of self-aligned superconducting cantilever probes with the patterning of weak links using a FIB to create nanoscale wireframe SQUID on cantilever probes. Nanowire frame tips with either open loops or sharp apices attached to pre-defined electrodes on a cantilever are made using corner lithography and standard microfabrication processes. Self-aligned superconducting probes are created by a shadow effect deposition of Nb by magnetron sputtering. A single step of FIB milling is used to create superconducting weak links at the apex of the tip, forming a SQUID-on-tip. The loop size of the SQUIDs can be controllably varied, from several micrometers down to 100 nanometers. The SQUIDs can be operated in magnetic fields up to and exceeding 1 T, and using the nanowire electrodes on-tip flux modulation can be supplied. Our approach shows the possibility to prepare and create versatile scanning probes based on superconducting devices on the wafer scale.\newline

\begin{figure*}[t]
    \centering
    \includegraphics[width=\linewidth]{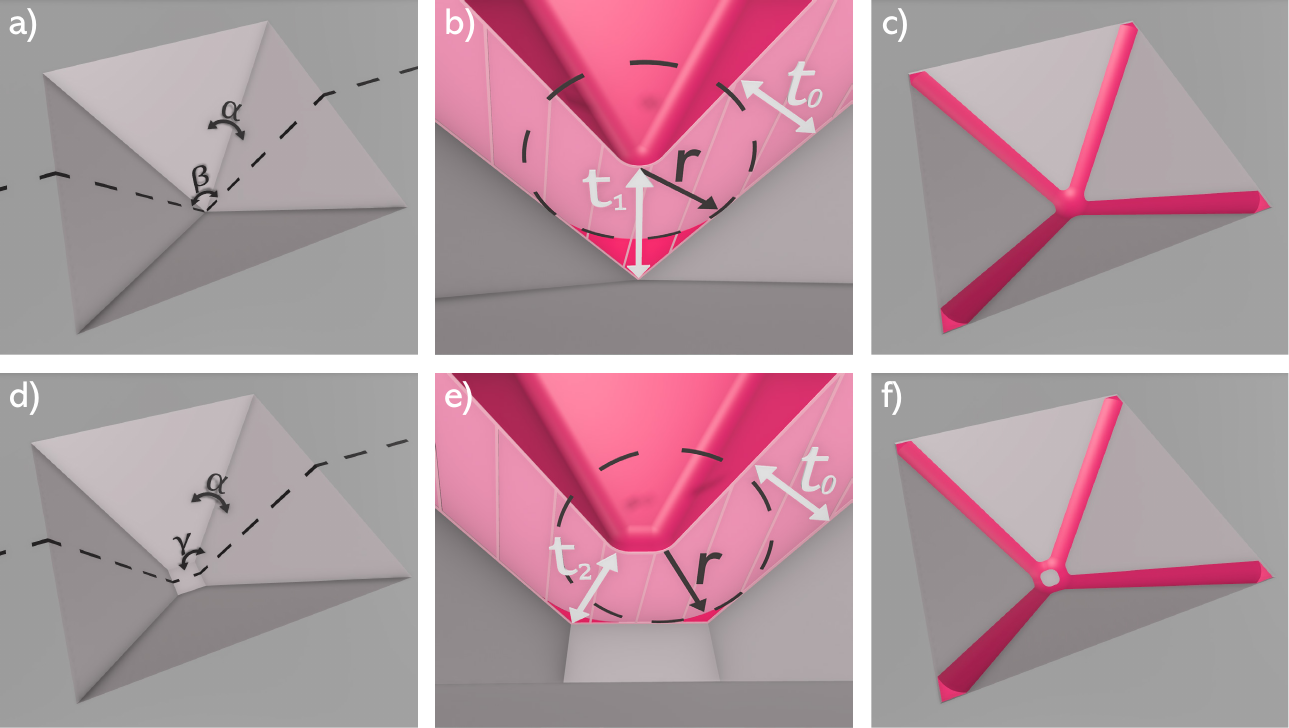}
    \caption{Schematic illustration of the principle of corner lithography. (a)/(d) Template preparation consisting of pits with concave corners $\alpha$, $\beta$, and $\gamma$. (b)/(e) Zoomed-in image of the conformal layer deposition (cross-section) at the pit bottom of (a)/(d). The thickness in the corners ($t_1$ and $t_2$) exceed the conformal layer thickness ($t_0$). After corner lithography, with an isotropic thinning distance $r$, nanostructures are formed in the corners of the template. (c)/(f) Corner lithography is able to make complicated three-dimensional structures by combining the nanostructures formed in different corners of the template with each other. The top (a)/(b)/(c) and bottom (d)/(e)/(f) row highlight that a variation in the initial template results in different nanostructures after the corner lithography process.}
    \label{fig:Fig1_CL}
\end{figure*}
Corner lithography is a versatile nanofabrication technique and has been used for wafer-scale fabrication of complex three-dimensional structures\cite{sarajlic_fabrication_2005, sarajlic_batch_2010, burouni_pyramidal_2012, berenschot_3d_2012, burouni_3d_2012,berenschot_fabrication_2013, burouni_wafer-scale_2013, pordeli_wafer-scale_2019, berenschot_self-aligned_2022, kooijman_lateral_2023, pordeli_vertical_2024, jonker_electrochemical_2024}. The technique of corner lithography is schematically illustrated in \figref{fig:Fig1_CL}{}. It consists of three basic steps: First, a template is created that contains concave corners, such as the inverted pyramid in \figref{fig:Fig1_CL}{a} and \figref{fig:Fig1_CL}{d}, with corners of angles $\alpha$, $\beta$, and $\gamma$. Second, a conformal deposition of material is performed that follows the geometry of the underlying template. The deposited layer will be thicker than the conformal layer thickness ($t_0$) in the corners of the template as can be seen in \figref{fig:Fig1_CL}{b} and \figref{fig:Fig1_CL}{e}. Lastly, the conformal layer is selectively etched for an isotropic thinning distance $r$ using time-controlled isotropic etching. This forms nanostructures in the corners with a thickness given by $t_0/\sin(\theta/2)-r$, where $\theta$ is the angle of the corner (\figref{fig:Fig1_CL}{b} and \figref{fig:Fig1_CL}{e}). \newline
\begin{figure*}[t]
    \centering
    \includegraphics[width=\textwidth]{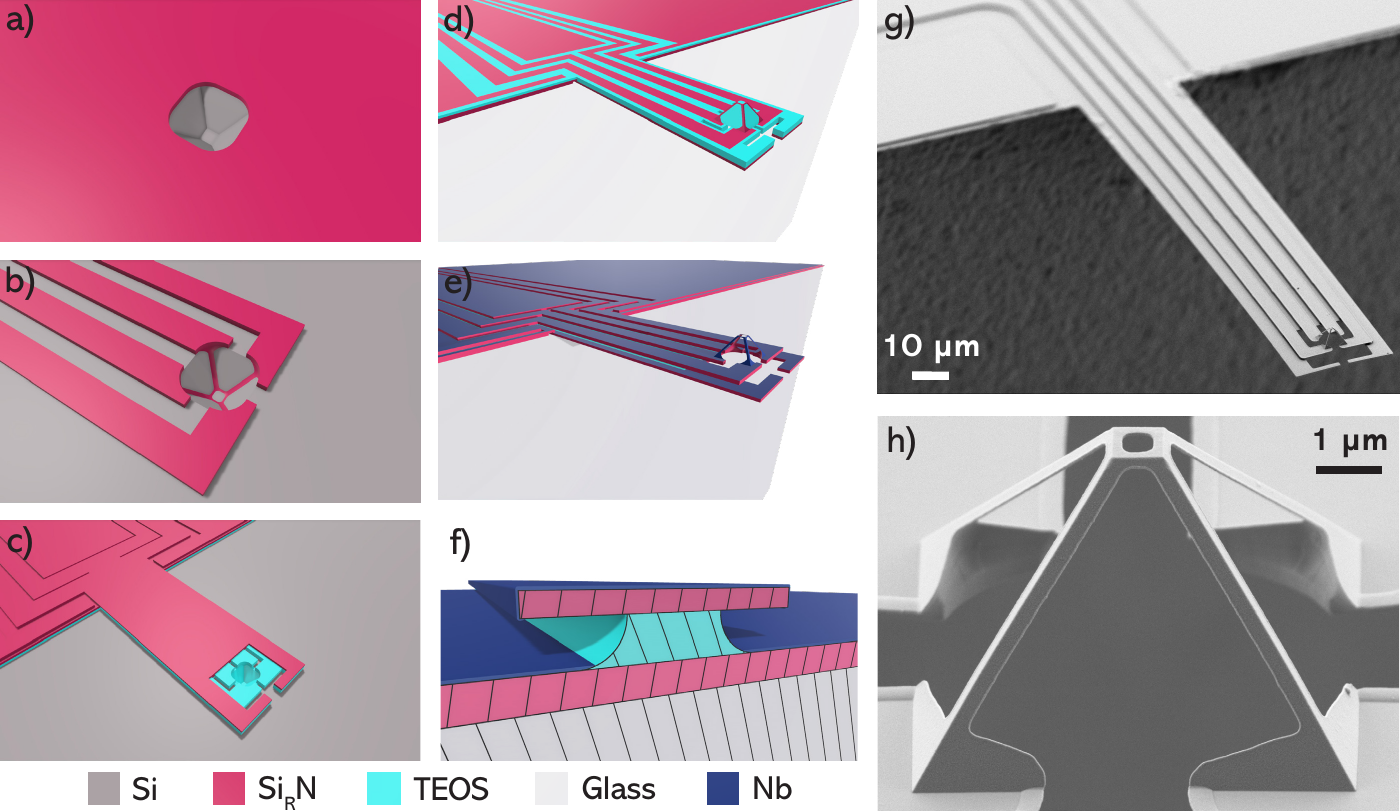}
    \caption{(a) Etched inverted truncated pyramid in Si (100), defined by a rounded \ce{Si_RN} mask. (b) Etching of electrodes in the \ce{Si_RN} mask, subsequent corner lithography is performed with an additional conformal layer of \ce{Si_RN} to create a wireframe with an open loop, which is connected to the electrodes. (c) Deposition of a heterostructure of TEOS/\ce{Si_RN}/TEOS (second TEOS layer not shown) and subsequent lithography and etching to define the cantilever and chip layout. (d) Anodic bonding of a diced glass wafer to the top TEOS layer, subsequent etching of the Si wafer in TMAH to release the freestanding cantilever probe (chip is flipped upside down with respect to (c)). (e) Magnetron sputtering of Ti/Nb/Pd/Au on the entire chip creates a self-aligned superconducting probe. (f) Cross-section image of the TEOS undercut and shadow effect deposition to electrically separate the \ce{Si_RN} chip and electrode layers. (g) Scanning electron microscopy (SEM) image of the created self-aligned superconducting cantilever probe with a tip at its end. (h) SEM image of the open-loop superconducting wireframe tip.}
    \label{fig:Fig2_Fab}
\end{figure*}
Corner lithography results in the formation of a nanowire for the angles $\alpha$ and $\gamma$ (formed by the intersection of two planes), whereas for the angle $\beta$ (created by the intersection of four planes at the apex of the inverted pyramid) a nanodot will be formed. By combining these different angles with one another, one can connect different nanofeatures together to create a self-aligned three-dimensional structure\cite{berenschot_3d_2012}. In \figref{fig:Fig1_CL}{a} the sharp inverted pyramid template results after corner lithography in a frame of nanowires that coalesces in a sharp dot as shown in \figref{fig:Fig1_CL}{c}. However, by only slightly adjusting the template to an inverted truncated pyramid (\figref{fig:Fig1_CL}{d}) the tip structure changes to a frame of nanowires connected to a square loop of nanowires, see \figref{fig:Fig1_CL}{f}. This shows the versatility that is offered by corner lithography to make complicated three-dimensional nanostructures by variations in the initial template. \newline

The process to fabricate the self-aligned superconducting wireframe on cantilever probes is shown in \figref{fig:Fig2_Fab}{}. First, templates for corner lithography are created in silicon (100) wafers. We deposit a 400 nm thick low-stress silicon-rich nitride (\ce{Si_RN}) layer on a Si (100) wafer using low pressure chemical vapor deposition (LPCVD). Square openings are defined in the \ce{Si_RN} by optical lithography and reactive-ion etching (RIE). To define templates for corner lithography, we use anisotropic etching of silicon in KOH, which results in the formation of inverted pyramids defined on the sides by the crystal (111) planes of the silicon. By controlling the etching time one can either make sharp inverted pyramids, where the four (111) planes intersect into a sharp apex (\figref{fig:Fig1_CL}{a}), or by prematurely stopping the etching one can create an inverted pyramid bound by a square bottom (100) plane (\figref{fig:Fig1_CL}{d} and \figref{fig:Fig2_Fab}{a}). Here, the dimensions of the initial mask opening and etching time define the size of the square bottom which can be varied from tens of nanometers to several micrometers. Next, optical lithography is used to define electrodes and contact pads in the \ce{Si_RN} mask, which connect to the four corners of the inverted pyramid. Subsequently, a conformal layer of \ce{Si_RN} is deposited using LPCVD. Corner lithography is performed on this 500 nm thick \ce{Si_RN} layer by time-controlled selective isotropic etching in hot \ce{H3PO4}. We overetch the \ce{Si_RN} layer by 5\% (overetch factor = 1.05), which ensures that corner lithography forms nanowires in the ribs of the inverted pyramid\cite{burouni_wafer-scale_2013}, nanowires in the ribs of the square bottom. And, if desired, a nanodot at the apex of the sharp inverted pyramid. Since the \ce{Si_RN} electrode layer overlaps with the inverted pyramid, the nanowire frame is physically connected to the electrodes, see \figref{fig:Fig2_Fab}{b} for the result. Next, we deposit a trilayer of tetraethyl orthosilicate (TEOS) (700 nm), \ce{Si_RN} (400 nm), and TEOS (110 nm) using LPCVD and the cantilever structure and chip layout are defined by optical lithography and RIE, see \figref{fig:Fig2_Fab}{c}. Using anodic bonding a glass wafer with pre-diced trenches is bonded to the silicon wafer, where we ensure the cantilevers are not covered by the glass wafer and instead extend over the trenches. In order to release the cantilever and corner-lithography-defined nanowire frame tip, the remaining glass above the cantilever is removed by dicing and consequently the entire silicon wafer is selectively etched using tetramethylammonium hydroxide (TMAH), see \figref{fig:Fig2_Fab}{d} for the resulting freestanding cantilever probe. The thick TEOS layer, which is sandwiched between the bottom \ce{Si_RN} cantilever/chip layer and top \ce{Si_RN} electrode layer, is selectively etched using buffered hydrofluoric acid (BHF) to create a physical undercut between the cantilever and electrode \ce{Si_RN} layers. This undercut creates a shadow effect between the two \ce{Si_RN} layers. The shadow effect will transfer to any superconducting layer deposited on the entire probe ensuring electrical separation of the device circuitry from the rest of the chip, see \figref{fig:Fig2_Fab}{e} for the superconducting cantilever probe and \figref{fig:Fig2_Fab}{f} for a zoomed-in cross-section image of the shadow effect deposition. Due to the large physical undercut it is also possible to use a less directional deposition technique, such as sputtering, to create a self-aligned superconducting probe. Conventional magnetron sputtering is then used to deposit a 5 nm adhesion layer of Ti, 60 nm layer of superconducting Nb, 2 nm capping layer of Pd, and finally a 20 nm thick layer of Au to serve as a protection layer during FIB. The final self-aligned superconducting cantilever probe is shown in \figref{fig:Fig2_Fab}{g}, which has contact pads and electrodes on the glass chip base, electrodes that extend onto a $\SI{150}{\micro m}$ long by $\SI{30}{\micro m}$ wide cantilever and are connected to a wireframe tip, see \figref{fig:Fig2_Fab}{h}. Using this technique wireframe tips with open loop sizes of several micrometers down to 300 nanometers have been made on the wafer scale, see Fig. S1 in the supplementary information. Moreover, by using multiple steps of corner lithography \cite{burouni_3d_2012, berenschot_fabrication_2013} or by using a thinner initial thickness of the conformal layer, the size of the created open loops can be reduced further.\newline
In order to show that the fabricated superconducting probes are sufficiently rigid and can also be used for scanning probe microscopy we have tested the probes in a room-temperature commercial AFM system. See Fig. S2 in the supplementary information for an AFM image taken in tapping mode of highly-oriented pyrolytic graphite (HOPG). The probe is able to resolve the step edges of the HOPG and does not deteriorate by the tapping mode scanning, thus showing its robust use for scanning probe microscopy.\newline

\begin{figure*}[t]
    \centering
    \includegraphics[width=0.7\linewidth]{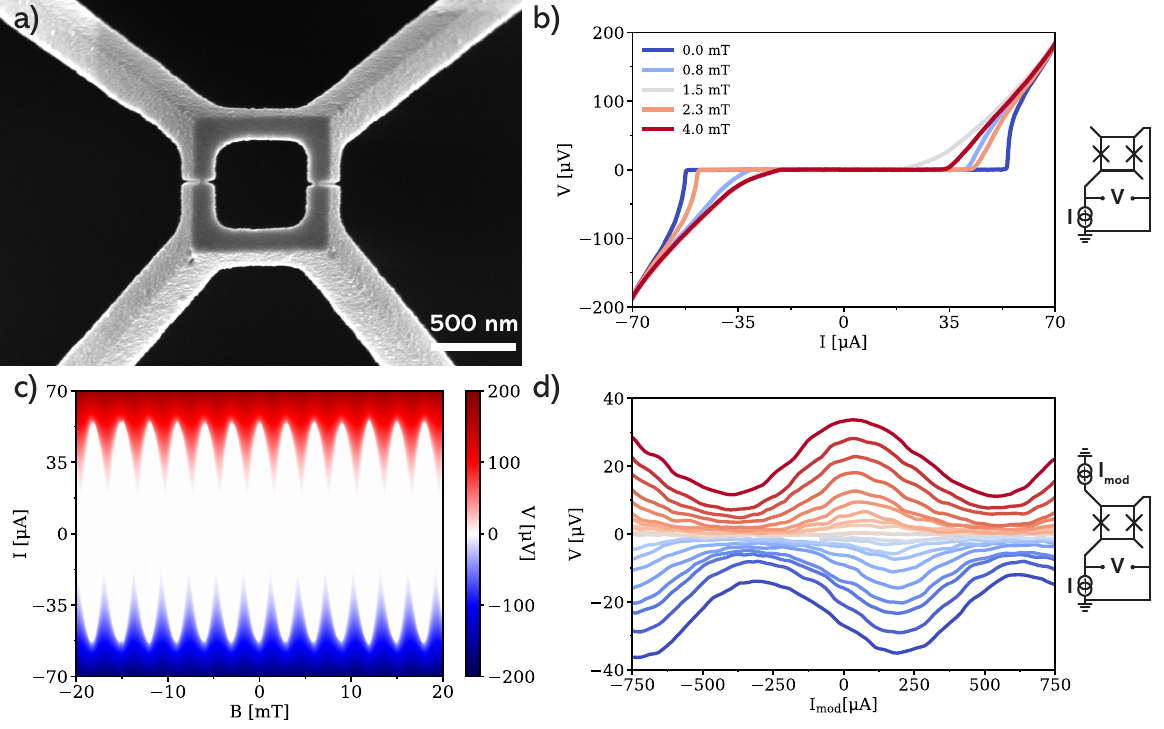}
    \caption{(a) Scanning electron microscopy image of a SQUID at the apex of an open-loop nanowire frame tip, the two Dayem bridges milled with FIB are also visible. (b) Current-voltage characteristics at 6.5 K for different applied fields, on the right a schematic is shown of the measurement circuit. (c) Quantum interference pattern of the SQUID. (d) On-tip flux modulation by a modulation current supplied with one of the nanowires in the wireframe, see the schematic on the right. The SQUID voltage oscillations can be seen for the applied modulation current for various bias currents ($\SI{-100}{\micro A}$ to $\SI{100}{\micro A}$ with steps of $\SI{10}{\micro\ampere}$ ) at T = 6 K.}
    \label{fig:Fig3_SQ}
\end{figure*}
\begin{figure*}[t]
    \centering
    \includegraphics[width=\linewidth]{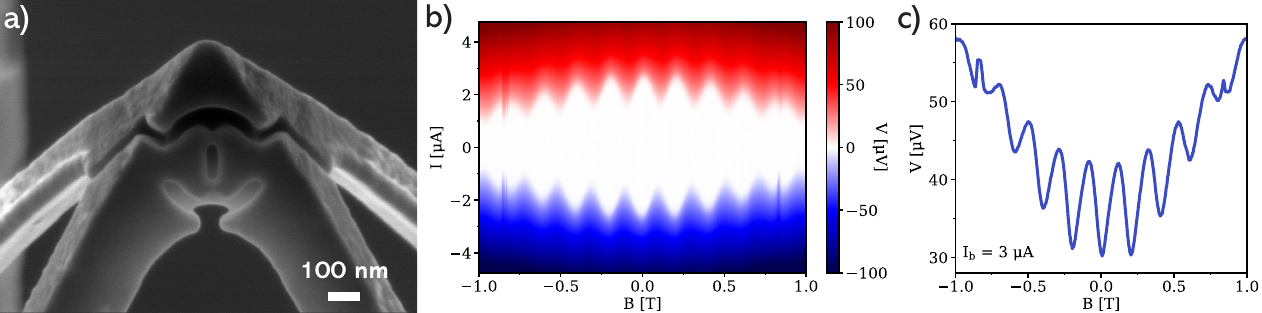}
    \caption{(a) Scanning electron microscopy image of a completely FIB-made nanoSQUID on a sharp nanowire frame tip. (b) Quantum interference pattern of a nanoSQUID with an effective diameter equal to 114 nm taken at a temperature of 3.25 K. The applied magnetic field is perpendicular to the cantilever probe. (c) SQUID oscillations in the normal state at $I_b = \SI{3}{\micro A}$ for a large range of magnetic field.}
    \label{fig:Fig4_nanoSQ}
\end{figure*}
In order to create a SQUID at the nanowire frame tip, FIB milling with Ga ions is used. This allows for the direct patterning of electrical circuitry at the tip of the probe. \figref{fig:Fig3_SQ}{(a)} shows constriction junctions in the nanowires of the open-loop wireframe milled by FIB. The FIB milling required to make the bridges is minimal, allowing the use of low beam power and short milling times. This reduces the amount of Ga which is implanted into the superconducting Nb, thus preserving the high critical temperature of Nb\cite{troeman_nanosquids_2007}. \figref{fig:Fig3_SQ}{(b)} shows the current-voltage characteristics of one of the FIB-milled SQUIDs. The I-V characteristics are measured using a simple current-bias scheme at a temperature of 6.5 K for different perpendicular magnetic fields. The I-V characteristics of the SQUID show no hysteresis at this operating temperature. However, when the temperature is decreased the device will exhibit a hysteretic dependence, which we attribute to Joule heating in the constriction junctions by high critical currents\cite{skocpol_selfheating_1974, blois_proximity_2013}. The I-V characteristics show a significant modulation of the critical current for an externally applied field, $I_c(B)$, of tens of $\SI{}{\micro A}$. \figref{fig:Fig3_SQ}{(c)} shows the quantum interference pattern associated to the I-V characteristics from \figref{fig:Fig3_SQ}{(b)}. The oscillation period of the pattern is 2.98 mT, which corresponds to an effective SQUID area of $\SI{0.69}{\micro m^2}$ and an effective square-loop SQUID size of 833 nm. Fig. S3 of the supplementary information shows three additional open-loop nanoSQUIDs with different effective areas. The smallest open-loop nanoSQUIDs that we have made have dimensions of 472 nm and a SQUID area of $\SI{0.22}{\micro m^2}$, this exceeds the geometrical hole area of the smallest open-loop wireframe (Fig. S1 supplementary information). We attribute this difference to flux focusing of the nanowires in the open-loop wireframe, which increases the effective area of the SQUID.\newline
From the modulation of the critical current $I_c(B)$ one can calculate the SQUID inductance parameter $\beta_{\text{L}}$ using $1/(1+\beta_{\text{L}}) = \Delta I_{\text{c}}/I_{\text{c,max}}$\cite{tesche_dc_1977}. We find $\beta_{\text{L}} = 0.53$ and from $\beta_{\text{L}} = 2LI_{\text{c}}/\Phi_0$ we estimate the SQUID inductance to be 20 pH, of which the main contribution originates from the kinetic inductance of the constriction junction bridges\cite{likharev_superconducting_1979, claassen_currentphase_1982, granata_nano_2016}. To characterize the sensitivity of the SQUID probes, we measure the flux noise spectral density of the SQUID at several applied magnetic fields, see Fig. S4 in the supplementary information. For a large range of frequencies, the lowest measured white flux noise floor was $\SI{3.8}{\micro\Phi_{\text{0}}/\sqrt{Hz}}$. This is consistent with SQUID systems lacking feedback loops or cryogenic amplification, where the noise profile is primarily limited by the room-temperature voltage amplification\cite{clarke_squid_2006, granata_nano_2016, martinez-perez_nanosquids_2017,normalNoise}. \newline
As indicated by the schematic in \figref{fig:Fig3_SQ}{b}, only two of the four nanowires in the wireframe are used for the current-bias readout. Hence, the remaining nanowires can be used to extend the functionalities of the on-tip circuitry. By using a shared ground, a modulation current $I_{\text{mod}}$ can be supplied using a nanowire electrode in addition to the bias current. As can be seen from the SQUID voltage oscillations in \figref{fig:Fig3_SQ}{d} the extra modulation current can be used to effectively flux-bias and modulate the wireframe SQUID. We can calculate the mutual inductance between the nanoSQUID and modulation line using, $M = \Phi_0/\Delta I_{\text{mod}}$. For the measured oscillation period of $\Delta I_{\text{mod}} = \SI{925}{\micro A}$ we find $M = \SI{2.23}{pH}$. However, note that due to the strong modulation current used, this also produces a significant stray field in close proximity to the tip. \newline
The use of on-tip flux-modulation allows for integrating the wireframe SQUID on cantilever together with a flux-locked loop to increase its sensitivity and operating range\cite{clarke_squid_2006, kirtley_fundamental_2010}. Additionally, the flexibility of FIB to mill an extra constriction bridge allows for operating the SQUID in a superconducting phase-locked loop\cite{uri_electrically_2016}. The on-tip circuitry could also be further extended by using the extra pair of nanowire electrodes for performing on-tip susceptometry, which is one of the main functionalities of SQUID-on-chip sensors\cite{kirtley_scanning_2016, persky_studying_2022}.\newline

As discussed earlier in \figref{fig:Fig1_CL}{}, slightly altering the initial template mold formed by KOH etching at the start of the fabrication changes the final tip structure. Changing from a KOH etch with a flat square bottom to one with a sharp point, like illustrated in \figref{fig:Fig1_CL}{(a)}, the final tip structure is also changed to a nanowire frame that coalesces in sharp dot, see \figref{fig:Fig1_CL}{(f)}. The FIB can be used in a similar fashion as in \figref{fig:Fig3_SQ}{(a)}, but this time to pattern the entire SQUID circuitry, including the nanoconstrictions, SQUID hole, trenches to reduce flux focusing and leads to connect the SQUID to the electrodes from the wireframe. In that case, the SQUID dimensions that can be realized are not limited to the waferscale fabrication process but to the resolution of the FIB. Hence smaller nanoSQUIDs can be made like the one shown in \figref{fig:Fig4_nanoSQ}{(a)}. \figref{fig:Fig4_nanoSQ}{(b)} shows the nanoSQUID quantum interference pattern with an oscillation period of 203 mT, which corresponds to an effective SQUID area of $\SI{0.01}{\micro m^2}$ and diameter of 114 nm. The pronounced $I_c(B)$ oscillations survive up to high fields, even above 1 T. In addition, the voltage oscillation at a given bias current, \figref{fig:Fig4_nanoSQ}{(c)}, also exists over a wide range of magnetic fields. Also note the small jumps both in the $I_\text{c}(B)$ in \figref{fig:Fig4_nanoSQ}{b} and the voltage oscillations in \figref{fig:Fig4_nanoSQ}{c} at $\pm\SI{0.8}{T}$. We attribute these jumps to the entering of a vortex in the wireframe loop and coupling into the nanoSQUID\cite{Lam_2011}. In addition, because the sharp nanowire frame tip is molded in the silicon crystal, the angles of the nanowires with respect to the cantilever plane are well defined at $\SI{54.3}{^o}$. Hence, a FIB-milled SQUID at the side of the tip as shown in \figref{fig:Fig4_nanoSQ}{a} is sensitive to both the in- and out-of-plane components of the magnetic field. This significantly increases the sensitivity of the probe to field sources that mainly produce out-of-plane magnetic fields like current lines\cite{persky_studying_2022}. Thus the wireframe nanoSQUID probe can be easily altered to fit the specific experiment to be performed.\newline

In summary, SQUIDs at the apex of wireframe tips defined by corner lithography have been created using focused ion beam milling. These SQUID on tips are integrated into self-aligned cantilever probes with on-chip device circuitry realized with shadow effect deposition on the wafer scale. The tip structure created by corner lithography can be easily modified by varying the initial mold. The versatility offered by both FIB and corner lithography allows for the fabrication of a wide range of SQUIDs with diameters ranging from several micrometers down to 100 nanometers, which can be operated in large external magnetic fields. Furthermore, since the wireframe tip is composed of four individual nanowires one can use the two of the nanowire electrodes to provide on-tip modulation of the SQUID. The bottom-up wafer-scale fabrication process to create the wireframe SQUID on cantilever probe enables integration of future on-chip sensing of the cantilever feedback during scanning, which will make the SQUID on cantilever a suitable probe for scanning probe microscopy in extreme environments, where optical AFM feedback is challenging.
\begin{suppinfo}
The supplementary information of this letter contains: i) SEM images of wireframe tips, ii) room-temperature AFM measurements, iii) additional nanoSQUID measurements, and iv) Flux noise measurements.
\end{suppinfo}

\begin{acknowledgement}
The authors acknowledge the research program “Materials for the
Quantum Age” (QuMat) for financial support. This program
(Registration No. 024.005.006) is part of the Gravitation
program financed by the Dutch Ministry of Education, Culture and Science (OCW). The authors would like to thank S. Haartsen and H.J.W. Zandvliet for their assistance in the room-temperature AFM measurements and S. Smink for useful discussions.
\end{acknowledgement}

\bibliography{20251104_Paper_SQUIDCantilever_v1}

\end{document}